\documentclass[useAMS,usenatbib]{mn2e}
\usepackage[pdftex]{graphicx} 
\usepackage[hidelinks]{hyperref}
\usepackage{xcolor} 
\usepackage{titlesec}
\usepackage{lscape}
\usepackage{calc}
\usepackage{amsmath}
\usepackage{afterpage}
\usepackage{bm}

\titleformat{\paragraph}
{\it\normalsize}{\theparagraph}{1em}{}
\titlespacing*{\paragraph}
{10pt}{3.25ex plus 1ex minus .2ex}{1.5ex plus .2ex}

\hypersetup{colorlinks,linkcolor={red!50!black}, citecolor={blue!50!black},urlcolor={blue!80!black} } 

\title[Thermal energy contents of the IGM with the SZ effect]
{Exploring the thermal energy contents of the intergalactic medium with
the Sunyaev-Zel'dovich effect}

\author[S.H. Lim et al.]{S.H. Lim$^{1}$\thanks{E-mail:
slim@astro.umass.edu}, 
H.J. Mo$^{1,2}$, 
Huiyuan Wang$^{3}$,
Xiaohu Yang$^{4}$
\\ \\
$^{1}$Department of Astronomy, University of Massachusetts, Amherst MA 01003-9305, USA \\
$^{2}$Physics Department and Center for Astrophysics, Tsinghua University, Beijing 10084, China \\ 
$^{3}$Key Laboratory for Research in Galaxies and Cosmology, Department of Astronomy,\\
~~University of Science and Technology of China, Hefei, Anhui 230026, China; \\
~School of Astronomy and Space Science, University of 
Science and Technology of China, Hefei 230026, China\\
$^{4}$Department of Astronomy, Shanghai Key Laboratory for Particle Physics and Cosmology, \\
~~Shanghai Jiao Tong University, Shanghai 200240, China; \\
~IFSA Collaborative Innovation Center, and Tsung-Dao Lee Institute, 
Shanghai Jiao Tong University, Shanghai 200240, China \\ 
} 

\begin{document} 

\date{Accepted ........ Received .......; in original form ......}

\pagerange{\pageref{firstpage}--\pageref{lastpage}}

\pubyear{2018}

\maketitle

\label{firstpage}

\begin{abstract} 
We examine the thermal energy contents of the intergalactic 
medium (IGM) over three orders of magnitude in 
both mass density and gas temperature using thermal 
Sunyaev-Zel'dovich effect (tSZE). The analysis is 
based on {\it Planck} tSZE map and the cosmic density field,  
reconstructed for the SDSS DR7 volume and sampled 
on a grid of cubic cells of $(1h^{-1}{\rm Mpc})^3$,  
together with a matched filter technique employed to 
maximize the signal-to-noise. 
Our results show that the pressure - density relation of the  
IGM is roughly a power law given by an adiabatic equation 
of state, with an indication of steepening at densities higher 
than about $10$ times the mean density of the universe.
The implied average gas temperature is $\sim 10^4\,{\rm K}$ in regions 
of mean density, $\rho_{\rm m} \sim {\overline\rho}_{\rm m}$,
increasing to about $10^5\,{\rm K}$ for 
$\rho_{\rm m} \sim 10\,{\overline\rho}_{\rm m}$, and to 
$>10^{6}\,{\rm K}$ for  $\rho_{\rm m} \sim 100\,{\overline\rho}_{\rm m}$. 
At a given density, the thermal energy content of the IGM is also 
found to be higher in regions of stronger tidal fields, likely 
due to shock heating by the formation of large scale structure and/or 
feedback from galaxies and AGNs.  
A comparison of the results with hydrodynamic simulations
suggests that the current data can already provide interesting 
constraints on galaxy formation.
\end{abstract} 


\section[intro]{INTRODUCTION}
\label{sec_intro}

The baryon mass that has been identified by current 
observations in galaxies and galaxy systems in the 
$z\sim 0$ universe accounts for about $2\%$ of the critical density 
\citep[e.g.][]{fukugita98, fukugita04}, 
which is about one third of the cosmic baryon abundance 
predicted by the primordial nucleosynthesis \citep[e.g.][]{olive00} 
and required to explain the fluctuations in the cosmic microwave 
background radiation (CMB) \citep[e.g.][]{wmap9, pcxiii}. 
Thus, a large fraction of baryons must be 
contained in the intergalactic medium. Indeed, 
observations of the Ly$\alpha$ forest at low-$z$ suggest
that about $30-40\%$ of the cosmic baryons may be hidden 
in the forest \citep[e.g.][]{danforth08}, but 
the results are still very uncertain, because the sample 
of low-$z$ forest is small and because of uncertainties 
in the ionization correction that is needed to obtain the 
total hydrogen density.

Cosmological simulations have shown that the cosmic baryons 
in the low-$z$ universe can exist in a variety of forms. 
In addition to stars and cold gas that are associated with 
galaxies, some baryons are predicted to be contained in 
the hot gaseous halos that are produced by the collapse 
of dark matter halos and feedback of galaxy formation. Furthermore, 
many gas simulations demonstrated that a large fraction 
of the baryons in the low-$z$ universe actually reside 
in a diffuse warm-hot intergalactic medium (WHIM) 
\citep[e.g.][]{cen99, dave99, dave01}, 
with temperature in the range between $10^5$ and $10^7\,{\rm K}$. 
This medium is generated by a combination of feedback 
from galaxy formation and shocks accompanying the formation of the 
large-scale structure 
\citep[e.g.][]{sunyaev72, nath01, furlanetto04, rasera06}. 

 Clearly, a comprehensive investigation of all 
gas components is required to have a complete understanding of 
how galaxies and larger scale structure form and evolve
in the universe. In particular, the study of 
the WHIM, which is potentially the dominating gas component in 
the low-$z$ universe but yet poorly understood so far, is important 
not only for obtaining a complete census of the cosmic baryons, 
but also for understanding how galaxies and structure formation 
interact with the gas component in the universe. Indeed, 
observations of Ly$\alpha$ absorption systems  
have revealed that the intergalactic medium (IGM) at $z\sim 0$
is significantly enriched in metal, indicating that a significant 
fraction of baryons may have been ejected from galaxies and
moved to the IGM by dynamical processes, such as winds driven 
by stellar and AGN feedback or by ram-pressure stripping 
\citep[e.g.][]{aguirre01}. In addition, the similarity 
in the metallicities of the intra-cluster media (ICM)
observed for different clusters suggests that the 
formation of the ICM is perhaps dominated by inflow of the 
gas from the large scale cosmic web 
\citep[e.g.][]{werner13, ettori15, mcdonald16, mantz17}.
Both of these demonstrate the importance of the interaction 
between galaxies and gas in shaping the gas media we 
observe. However, the detection of the WHIM in observations is 
challenging. At a temperature of $T\sim 10^{5}$-$10^{7}\,{\rm K}$, 
the gas is almost completely ionized, making it difficult 
to detect in absorption. The low density of the WHIM also makes 
it difficult to detect in UV and X-ray emission.

With the advent of large surveys of the cosmic microwave background
(CMB), a promising new avenue is opened up. As the CMB 
photons travel to reach us, they are scattered by free electrons 
associated with hot gas via Compton scattering. This 
produces a change in the energy distribution of the CMB photons, 
an effect referred to as the thermal Sunyaev-Zel'dovich 
effect \citep[tSZE;][]{sunyaev72}. The tSZE is a measure of 
the projected electron pressure along the line of sight, and 
so it provides a way to probe the thermal energy content of 
the ionized gas in the universe. Compared to X-ray emissions, 
the tSZE is more sensitive to diffuse gas, making it more 
suitable for probing relatively low-density media, such as 
WHIM. It also complements absorption studies of highly 
ionized gas, as ionization correction and gas metallicity 
are not needed to obtain the total gas content.  

 Great amounts of effort have been made to measure the tSZE 
from observational data. \citet{pcxi} used 
the {\it Planck} multi-frequency CMB temperature maps and 
a sample of locally brightest galaxies to study the 
tSZE produced by galaxy systems down to a halo mass 
$\sim 4\times 10^{12}\,{\rm M_\odot}$. Remarkably, they found that 
their results are consistent with the self-similar model in which 
the hot gas fraction in halos is independent of halo mass. 
A similar conclusion was also reached by \citet{greco15} using 
a similar method. \citet{tanimura17} and \citet{degraaff17} 
reported the detection of warm-hot gas in cosmic filaments 
by cross-correlating filaments identified by galaxy pairs 
with the {\it Planck} tSZE map. By cross correlating 
the {\it Planck} tSZE map with the mass density map obtained 
from the gravitational lensing data of the CFHTLenS survey, 
\citet{vanwaerbeke14, ma15, hojjati15} 
found significant correlations 
between the gas and dark matter distributions. Finally, 
\citet{hill14} reported a significant detection 
of cross-correlation between the {\it Planck} CMB lensing 
potential map and the {\it Planck} tSZE map. All these 
indicate that tSZE provides a powerful way to investigate 
the gas distribution in the universe. 

Recently, \citet{lim18a, lim18b} used the group catalogs of 
\citet{yang07} and \citet{lim17} to extract, 
from the {\it Planck} CMB observation, both the tSZE and 
kinematic SZ effects (kSZE) of the warm-hot gas associated with 
halos of different masses. They employed the matched filter technique 
\citep{haehnelt96, herranz02, melin05, melin06} 
to maximize the signal-to-noise ratio. In particular, they matched 
the filter simultaneously to all galaxy groups to constrain 
the corresponding signals jointly as a function of group mass, 
so as to minimize projection effects for groups lying closely 
in the sky. Combining the tSZE and kSZE, they found that the total 
amount of the baryons associated with dark matter halos are 
consistent with the universal baryon fraction, even in low-mass halos, 
but that the gas temperature is much lower than the virial 
temperature in low-mass halos. Various tests performed demonstrate 
that the method is very powerful in extracting the SZE signals 
effectively and reliably. 

In this paper, we study the thermal contents of the IGM at $z\sim0$, 
extending the analyses of \citet{lim18a, lim18b} by going beyond 
scales of dark matter halos. Specifically, we constrain the pressure 
of the gas in different environments, using the total density field 
reconstructed by \citet{wang09, wang16} in the Sloan Digital Sky Survey
Data Release 7 (SDSS DR7) volume and using the {\it Planck} tSZE map 
as the observational constraint. As we will see, our analysis is 
able to provide constraints on the thermal energy contents 
of the WHIM in regions covering almost three orders of magnitude
in mass density. In particular, our results are shown to be 
capable of providing constraints on models of galaxy formation. 
The outline of this paper is as follows. We describe the  
data used in our analysis in Section~\ref{sec_data}, and our method to 
constrain the gas pressure in Section~\ref{sec_methods}. 
We present our main results and comparisons with results 
from numerical simulations in Section~\ref{sec_results}. 
We summarize and conclude in Section~\ref{sec_sum}.

\section[data]{DATA}
\label{sec_data}

\subsection{The \textit{Planck} \textit{y}-map} 
\label{ssec_ymap}

The temperature change in the CMB spectrum by the tSZE is given by
\begin{eqnarray}\label{eq1}
\left(\frac{\Delta T}{T_{\rm CMB}}\right) = 
g(x) y \equiv 
g(x)\frac{\sigma_{\rm T}}{m_{\rm e}c^2} \int{P_{\rm e} {\rm d}l}, 
\end{eqnarray}
where $y$ is the Compton parameter, $g(x)=x \coth(x/2)-4$ 
is the conversion factor at a given 
$x\equiv h\nu/(k_{\rm B}T_{\rm CMB}$), $T_{\rm CMB}=2.7255\,{\rm K}$, $\sigma_{\rm T}$ is 
the Thompson cross-section, $m_{\rm e}$ is the electron rest-mass, 
$c$ is the speed of light, $P_{\rm e}=n_{\rm e}k_{\rm B}T_{\rm e}$ is the electron pressure
with $n_{\rm e}$ the free electron density, and ${\rm d}l$ is the path length along 
each line-of-sight (LOS). 
The {\it Planck} \citep{tauber10, pci} is an all-sky observation of 
the CMB in nine frequency bands from $30$ to $857$ GHz, 
with angular resolutions ranging between $5\arcmin$ and $31\arcmin$. 
We use the {\it Planck} MILCA \citep[Modified Internal Linear 
Combination Algorithm;][]{hurier13} all-sky tSZ Compton parameter
map \citep{pcxxii}, also known as the MILCA $y$-map, 
which is part of the {\it Planck} 2015 data release 
\footnote{\url{https://pla.esac.esa.int}}.
The MILCA $y$-map is constructed from the full mission data set of the 
{\it Planck}, using a combination of different frequency maps to minimize the 
primary CMB fluctuations and the contamination from foreground sources. 
The details about the $y$-map construction can be found in the original papers. 
We mask the brightest $40\%$ of the sky to limit the Galactic foreground 
contamination, by using the corresponding mask provided in the {\it Planck} 
2015 data release. We also apply the mask for point sources, provided in the 
same data release, to reduce the contamination from radio and infrared 
sources. 

As a test, we have also applied the same analysis to the NILC 
\citep[Needlet Independent Linear Combination;][]{remazeilles11} 
$y$-map \citep{pcxxii}, which treats dust contamination 
differently than the MILCA. No significant change is found 
in our results.

\subsection{The dark matter density field in the SDSS DR7 volume} 
\label{ssec_den}

Another set of data that we use for our analysis is the reconstructed 
cosmic density field in Sloan Digital Sky Survey Data Release 7 
\citep[SDSS DR7;][]{abazajian09} volume given by \citet{wang16} (W16). 
The reconstruction uses galaxy groups selected with the 
halo-based group finder \citep{yang05, yang07} to represent 
dark matter halos. Extensive tests using mock galaxy redshift surveys 
constructed from the conditional luminosity function (CLF) model 
\citep[e.g.][] {yang03, vdB03} and semi-analytical models \citep[][]{kang05} 
revealed that this group finder is very successful in grouping galaxies 
into their common dark matter halos \citep[see][]{yang07}. 
By partitioning the SDSS volume into domains associated with 
individual groups, and by modelling the mass distribution
in each domain using profiles calibrated with $N$-body simulations, 
W16 reconstructed the real space density field within the entire SDSS DR7 
volume. W16 used groups of halo masses 
$M_{\rm h}\geq 10^{12}h^{-1}{\rm M_\odot}$, so that  
completeness can be achieved to $z\sim 0.12$. The 
reconstruction was restricted to the contiguous region of the 
SDSS DR7 in the Northern Galactic Cap of 
$\sim 7,000\,{\rm deg}^2$. While W16 provides the density field 
smoothed on various scales, we use the density field smoothed on 
$1\, h^{-1}{\rm Mpc} $ for our analysis. 

\subsection{Hydrodynamic simulations for comparison} 
\label{ssec_simul}

\begin{figure}
\includegraphics[width=1.\linewidth]{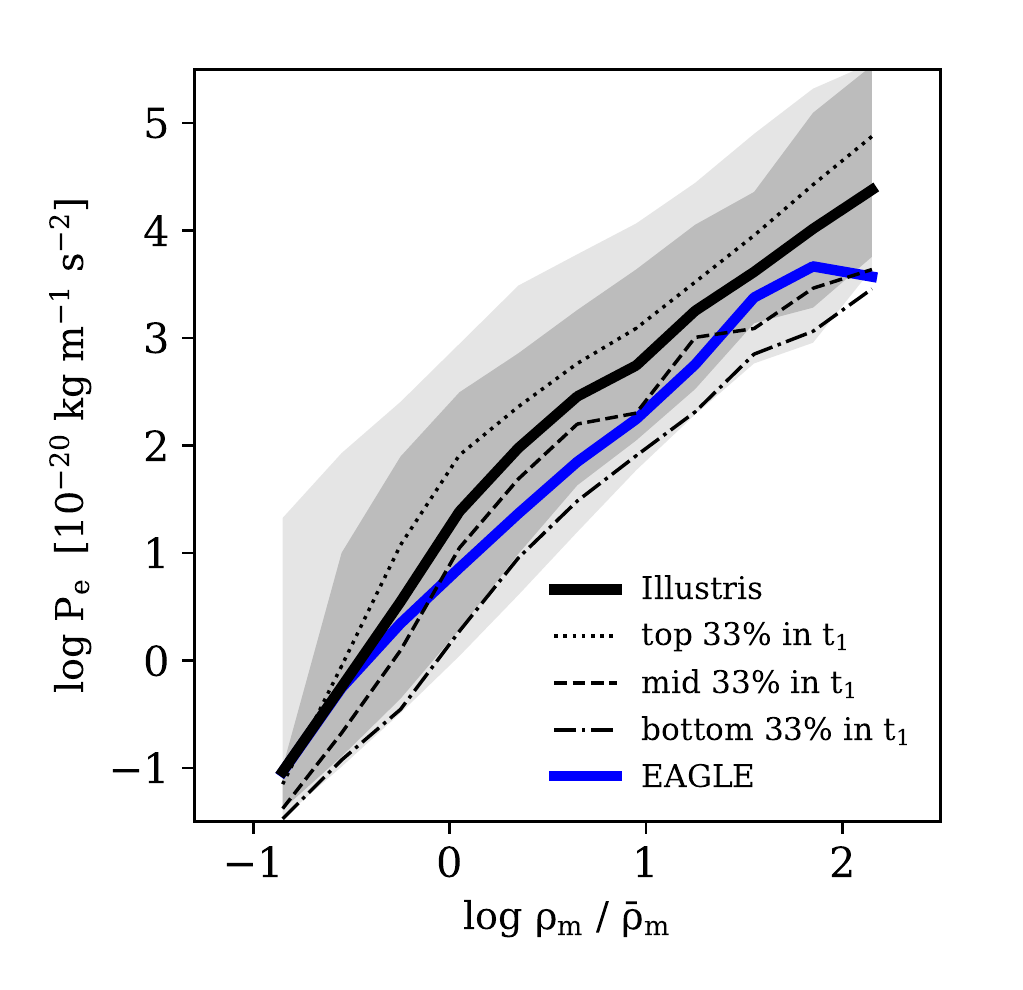}
\caption{The relations between the electron pressure and 
mass density for grid-cells of 
$1\, (h^{-1}{\rm Mpc})^3$ obtained from Illustris and EAGLE. 
The solid lines show the mean relations, and the gray bands show the $1\sigma$ 
and $2\sigma$ dispersion obtained from Illustris. The thin lines show 
the mean relations obtained for the three sub-samples of the grid-cells 
according to the ranking in the tidal field strength, $t_1$, 
at given density, obtained from Illustris.}
\label{fig_sim}
\end{figure}

\begin{figure}
\includegraphics[width=1.\linewidth]{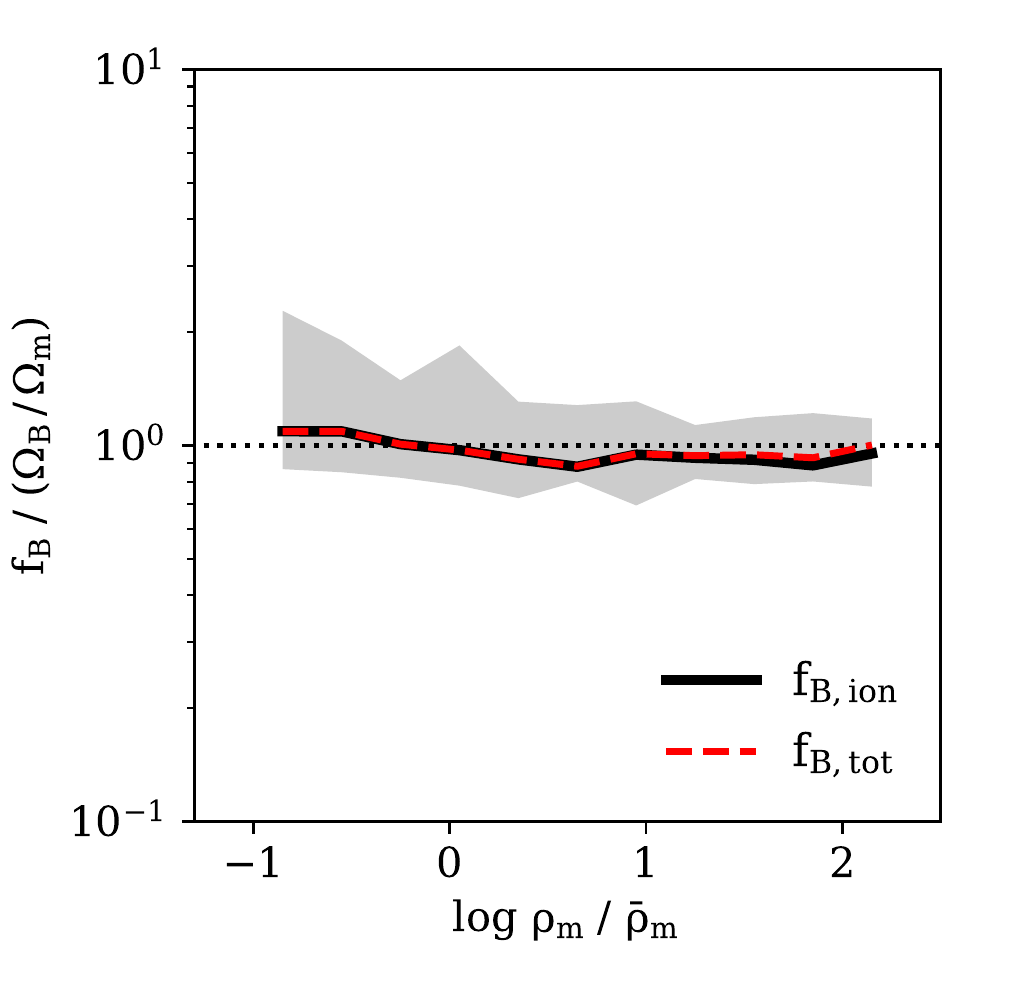}
\caption{The ionized gas mass fraction, normalized by  
the cosmic mean baryon fraction, in grid-cells as a function 
of mass density as obtained from 
Illustris. The thick line shows the mean relation, while the shaded 
region indicates the $1\sigma$ dispersion. For comparison 
the dashed line shows the average total baryon fraction as a 
function of mass density.} 
\label{fig_fB_ion}
\end{figure}

 To compare with our results, we use two recent cosmological 
hydrodynamic simulations. The first is Illustris 
\citep[][]{nelson15}, which follows  
the evolution of the simulated Universe in a box of 
$(106.5\,{\rm Mpc)^3}$, implementing physical processes such 
as radiative cooling, star formation, and various feedback processes. 
The free parameters in sub-grid models were constrained by 
observations as well as simulations of higher resolutions. 
We use Illustris-1, their flagship run, which has the highest 
mass resolution, with gas particle mass of $1.6\times10^6\,{\rm M_\odot}$. 
The simulation adopted WMAP9 cosmology, 
with $\Omega_{\rm m}=0.273$, $\Omega_\Lambda=0.727$, 
$h=0.704$, and a baryon fraction, 
$\Omega_{\rm B}/\Omega_{\rm m}=0.165$  
\citep{wmap9}. 

Another simulation we use is 
the Evolution and Assembly of GaLaxies and their Environments 
\citep[EAGLE;][]{schaye15, crain15, mcalpine15}.
EAGLE implemented sub-grid physics models for cooling, star formation, and 
stellar and AGN feedback to evolve different types of particles, such as gas, 
stars, dark matter, and black holes. The parameters for these models are tuned 
to match a set of observations, such as the stellar mass function and stellar 
mass - black hole mass relation at $z\sim0$. EAGLE provides a 
number of simulations assuming different model parameters and box sizes. 
We present results based on the `RefL0100N1504' run, their 
fiducial simulation, which has the largest box size,  
$(100\,{\rm Mpc)^3}$, among all the EAGLE runs. 
Our tests showed that the results used for our comparisons 
were not significantly affected if other simulations were 
used. EAGLE assumes the {\it Planck} cosmology, 
$(\Omega_{\rm m}, \Omega_\Lambda, h)=$ 
$(0.307, 0.693, 0.678)$ \citep{pcxiii}. 

Fig.\,\ref{fig_sim} shows the pressure - density relation averaged 
over grid-cells of $1\, (h^{-1}{\rm Mpc})^3$ in the simulations.
The cell size chosen here is to match the size adopted in 
our analysis. The mean relations, presented by the thick solid lines, 
have slopes roughly consistent with the adiabatic equation of state, 
$P_{\rm e}\sim \rho^{5/3}$ for both simulations. However, 
the average pressure at $\rho_{\rm m}>{\overline\rho}_{\rm m}$
predicted by Illustris is about two to three times as high as 
that predicted by EAGLE, presumably because of cosmic variance 
and of the differences in the feedback models adopted 
in the two simulations. 
The scatter in pressure at a given density (the gray bands showing 
the $1\sigma$ and $2\sigma$ dispersion given by Illustris) 
is also very large. This large scatter motivates us to explore the 
possibility of a second parameter, in addition to density, 
that can affect the gas pressure (see \S\ref{ssec_tidal}).

On the scale of a grid cell, the distribution of baryons 
is expected to approximately follow that of the dark matter. 
Since most of the baryons are expected to be in the state
of diffuse ionized gas, the total amount of ionized gas in a 
grid cell is roughly that given by the universal baryon fraction. 
Fig.\,\ref{fig_fB_ion} shows that this is indeed the case, 
at least in simulations. This result will be used to 
interpret the tSZE obtained from the observational data.  

\subsection{The large-scale tidal field} 
\label{ssec_tidal}

We have tested a number of quantities available from the 
simulation to see how they affect the gas pressure. 
These include stellar mass, star formation rate, black hole mass, 
mean velocity, velocity dispersion, the fluctuation in 
dark matter density, distance to nearest halos, fraction
of mass contained in halos, and local tidal field, all 
estimated for individual grid cells that are used to sample the
density field. We found that, among these quantities,  
the large-scale tidal field is the most significant second 
parameter that can change the gas pressure on top of 
the dependence on local dark matter density. 

In our analysis, we tested two definitions of large-scale 
tidal field, one based on dark matter halos and 
the other based on the mass density field. 
We find that both definitions of the tidal field give 
practically the same results in our analysis 
[see \citet{wang11} for details of how the two 
tidal fields are correlated]. In what follows, we only 
present results using the halo-based tidal field.
The estimate of the halo-based tidal field is 
based on the halo tidal force along a direction ${\bf t}$ 
exerted on the surface of a sphere of diameter 
$1\, h^{-1}{\rm Mpc}$ that approximates a grid-cell, 
normalized by the self-gravity of the matter inside the grid-cell, 
\begin{eqnarray}
f({\bf t}) = \frac{\sum_{i}GM_iR_{\rm g} (1+3\cos 2\theta_i)/r_i^3}
{2GM_{\rm g}/R_{\rm g}^2}
\end{eqnarray}
where the summation is over all the halos, 
$M_i$ is the mass of halo $i$, $R_{\rm g}=0.5 h^{-1}{\rm Mpc}$ 
is the radius of the sphere that approximates the grid-cell, 
$M_{\rm g}$ is the mass enclosed within 
the grid-cell in question, 
$r_i$ is the separation between the center of the grid-cell  
and the $i$-th halo, and $\theta_i$ is the angle between 
${\bf t}$ and ${\bf r}_i$ \citep[e.g.][]{wang11}. 
So defined, the tidal field measures the total tidal 
force exerted on a grid-cell normalized by its self-gravity.  
The ellipsoid of the local tidal field is then used to compute the 
eigenvalues $t_1$, $t_2$, and $t_3$ ($t_1\geq t_2 \geq t_3$) of the 
halo tidal tensor. We use $t_1$ to describe the strength 
of the halo tidal field at any given grid-cell of 
$1\, (h^{-1}{\rm Mpc})^3$. 

Fig.\,\ref{fig_sim} shows that grid-cells with higher 
$t_1$ have higher mean pressure at a given density 
in the Illustris simulation, and the result for 
the EAGLE simulation is qualitatively the same. 
Apparently, the formation of large scale structure and/or 
intense feedback in regions of strong tidal field can heat 
the IGM on large scales. 

\section[methods]{METHODS}
\label{sec_methods}

\subsection{The matched filter technique}
\label{ssec_mf_gen}

Extracting the tSZE signals reliably requires to optimize the signal-to-noise,  
since the signal is generally more than an order of magnitude lower than other 
sources, such as the primary CMB anisotropy, Galactic foreground, 
and cosmic infrared background. Thus, using a simple aperture photometry 
can lead to large uncertainties in the results \citep[see e.g.][]{melin06}. 
Here, we employ the matched filter (MF) technique 
\citep{haehnelt96, herranz02, melin05, melin06}, 
which is designed to minimize source confusions and contamination,
and to maximize the signal-to-noise by imposing prior 
knowledge of the signals and the noise power spectra. In the MF technique, 
the Fourier transform of the filter that optimizes the signal-to-noise 
is given by
\begin{eqnarray}\label{eq_MF}
\hat{F}(\boldsymbol{k}) = \left[\int 
\frac{|\hat{\tau}(\boldsymbol{k'})\hat{B}(\boldsymbol{k'})|^2}
{P(k')} 
\frac{{\rm d}^2k'}{(2\pi)^2}  \right]^{-1} 
\frac{\hat{\tau}(\boldsymbol{k}) \hat{B}(\boldsymbol{k})}
{P(k)}
\end{eqnarray}
where, in our application, $\hat{\tau}(\boldsymbol{k})$ is 
the Fourier transform of the projected electron pressure, 
$\hat{B}(\boldsymbol{k})$ is the Fourier transform
of the Gaussian beam function that mimics the convolution in 
the {\it Planck} observation with the FWHM of $5\arcmin$, 
and $P(k)$ is the noise power spectra. Because the MILCA $y$-map is already 
cleaned of the primary CMB anisotropy, $P(k)=P_{\rm noise}$ where $P_{\rm noise}$ 
is the power spectrum of the noise map for the MILCA $y$-map, as provided in the 
{\it Planck} data release. Both the shape and amplitude of the 
projected electron pressure profile, $\hat{\tau}(\boldsymbol{k})$, are simultaneously 
constrained by matching the filters to all pixels, as to be described in 
details below. 

\subsection{The pressure - density relation}
\label{ssec_mf_app}

We assign a value of electron pressure to each of the grid-cells by assuming 
a simple double power-law relation between the reconstructed matter density 
field and the pressure: 
\begin{eqnarray}\label{eq_pl}
P_{\rm e} = 
\begin{cases} 
A\times(\rho_{\rm m}/\rho_{\rm m,0})^{\alpha_1}, 
& \mbox{if } \rho_{\rm m} \leq \rho_{\rm m,0} \\ 
A\times(\rho_{\rm m}/\rho_{\rm m,0})^{\alpha_2}, 
& \mbox{if } \rho_{\rm m} > \rho_{\rm m,0}. 
\end{cases}
\end{eqnarray}
The pressure field is smoothed with a Gaussian kernel of radius 
$1\, h^{-1}{\rm Mpc}$, and is integrated along each line of sight 
to obtain the predicted $y$-parameters for all the pixels in in the 
{\it Planck} $y$-map. The predicted $y$-parameter profile at each 
pixel is used in equation (\ref{eq_MF}) to obtain the corresponding 
filter, and the filters for all the pixels are matched to the 
{\it Planck} map. Finally, we perform the Monte Carlo Markov Chain 
(MCMC) to constrain the parameters of the double power-law 
relation so as to 
yield the best overall match between the filters and the observed 
$y$-map, based on the sum of the $\chi^2$ over all the pixels. 
We assume a constant background contribution from free electrons 
outside the volume in which the density reconstruction was made, 
and the background level is treated as a free parameter to be 
constrained with the MCMC. This is expected to be valid 
as long as the structures lying beyond the boundary of the 
reconstruction volume are not correlated with the structures 
within the volume. In this case, the fluctuations of the 
background do not lead to bias, but increase noise. 

\section[results]{RESULTS}
\label{sec_results}

\subsection{The pressure - density relation} 
\label{ssec_P_rho}

\begin{figure}
\includegraphics[width=1.\linewidth]{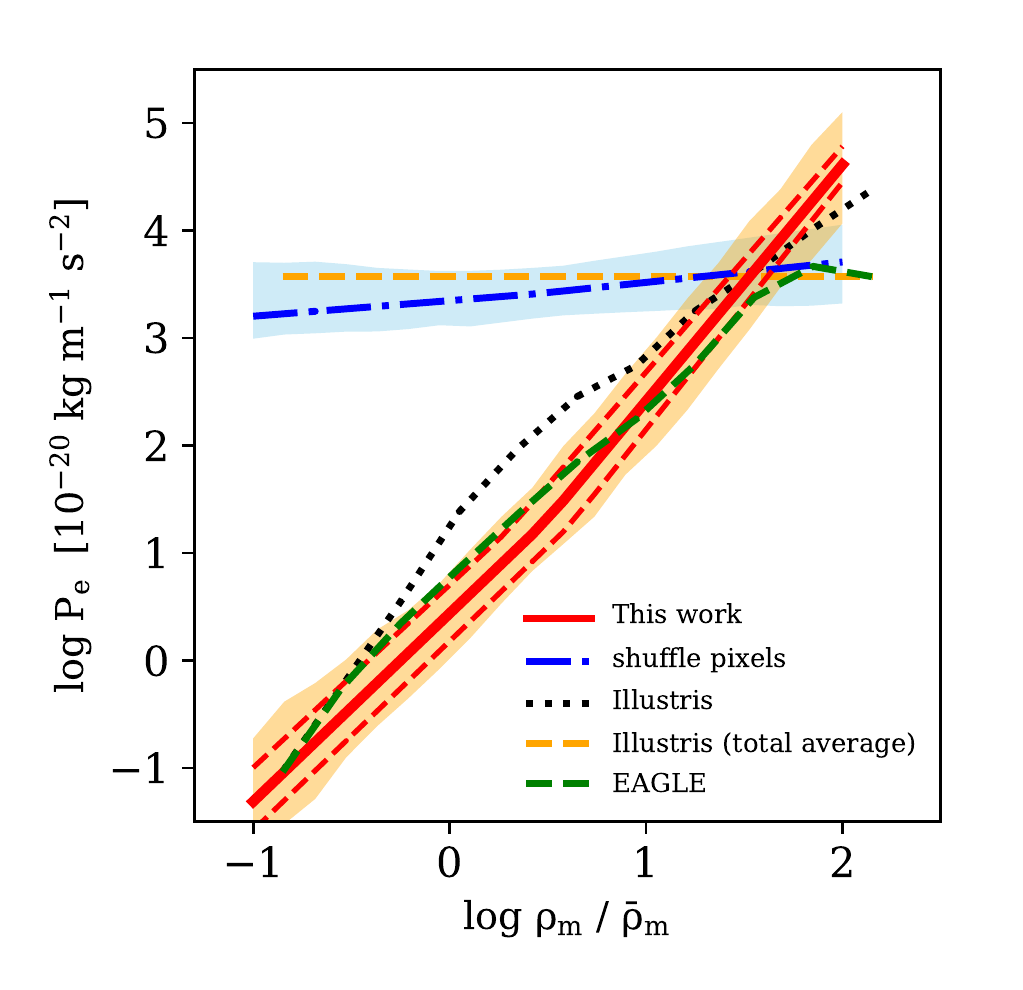}
\caption{The pressure - density relation, compared with the simulations. 
The red line shows the mean relation obtained with our method, with 
the orange band showing the $1\sigma$ dispersion estimated from the 
uncertainties in the constrained parameters, as given by the MCMC
sample. The red dashed lines are the relations constrained separately for 
the two nearly equal-sized sub-samples of the grid-cells according to 
the sky region, one including the Sloan Great Wall \citep{gott05}, 
and the other not. The black and green lines 
show the mean relations from Illustris and EAGLE, respectively. 
Also, the mean relation obtained from shuffling the grid-cells (dot-dashed), 
with the band showing the relations from $50$ realizations, is compared 
with that obtained from Illustris (orange dashed).}
\label{fig_P_rho}
\end{figure}

\begin{table}
 \renewcommand{\arraystretch}{1.5} 
 \centering
 \begin{minipage}{80mm}
  \caption{The constrained parameters for the power-law relation.}
  \begin{tabular}{ccccc}
\hline
Samples & $A$\textsuperscript{\footnote{The values are in 
the unit of $10^{-20}\,{\rm kg}\,{\rm m}^{-1}\,{\rm s}^{-2}$.}} & $\rho_{\rm m,0}\,/\,{\overline\rho}_{\rm m}$ & $\alpha_1$ & $\alpha_2$  \\
\hline
\hline
all cells & $20\pm6.9$ & $3.0\pm1.5$ & $1.7\pm0.28$ & $2.2\pm0.12$ \\ 
high $t_1$ & $25\pm26$ & $1.1\pm2.1$ & $1.9\pm0.48$ & $1.8\pm0.26$ \\
mid $t_1$ & $21\pm20$ & $2.1\pm3.1$ & $1.7\pm0.54$ & $1.9\pm0.20$ \\
low $t_1$ & $2.5\pm17$ & $1.4\pm2.5$ & $1.8\pm0.43$ & $2.2\pm0.24$ \\

\hline
\\
\vspace{-8mm}
\end{tabular}
\textbf{Notes.}
\vspace{-5mm}
\label{tab_param}
\end{minipage}
\vspace{5mm}
\end{table}

Fig.\,\ref{fig_P_rho} shows the pressure - density 
relation obtained from the data, with the constrained parameters listed in 
Table\,\ref{tab_param}. The orange band shows the $1\sigma$ 
scatter based on the uncertainties of the constrained 
parameters given by the MCMC, representing 
the dispersion in the relation. The slope of the relation 
for low-density region is very close to that expected from an 
adiabatic equation of state, $P_{\rm e}\sim \rho^{5/3}$, 
but the relation is slightly steeper for dense regions. This indicates 
that gas in the dense regions may be heated by additional 
sources, such as supernova and AGN feedback from galaxies 
residing in these regions, and shocks associated with the 
formation of large scale structure. As a null test, we randomly 
shuffle the grid-cells and apply the same method to constrain the 
relation. The mean relation, plotted as the thick blue line, 
and the relations from a total of $50$ realizations, 
plotted as the sky-blue band, are all flat, as 
expected from random fluctuations. This demonstrates that 
our method is able to detect the true relation between the 
gas and dark matter distribution. Note that the scatter 
in the constrained pressure - density relation is similar to
that among the random samples, indicating that the 
uncertainty in the constrained relation is dominated 
by fluctuations of the background/foreground. Finally, 
we divide the grid-cells into two nearly equal-sized 
sub-samples according to the sky region they belong to, one 
including the Sloan Great Wall \citep{gott05} and the other 
not, and constrain the relation separately for the two sub-samples. 
The mean relations obtained for the two sub-samples, 
shown by the red dashed lines, are within the $1\sigma$ scatter 
band obtained for the entire sample, indicating that  
cosmic variance does not affect our results significantly. 

We find that using a different smoothing scale, e.g. 
$2\, h^{-1}{\rm Mpc}$, leads to no significant change in our 
results. Choosing an even larger smoothing scale leads to
bigger uncertainties in the constrained relations, 
because of the decreased number of grid-cells. Note that the 
uncertainty in the reconstructed peculiar velocities is 
about $100\,{\rm km\,s^{-1}}$ (see W16), which corresponds 
to $\sim 1\, h^{-1}{\rm Mpc} $ in real-space positions.  
Thus, choosing a smoothing scale smaller than 
$\sim 1\, h^{-1}{\rm Mpc} $ may not be appropriate.

\subsection{The temperature - density relation} 
\label{ssec_T_rho}

\begin{figure}
\includegraphics[width=1.\linewidth]{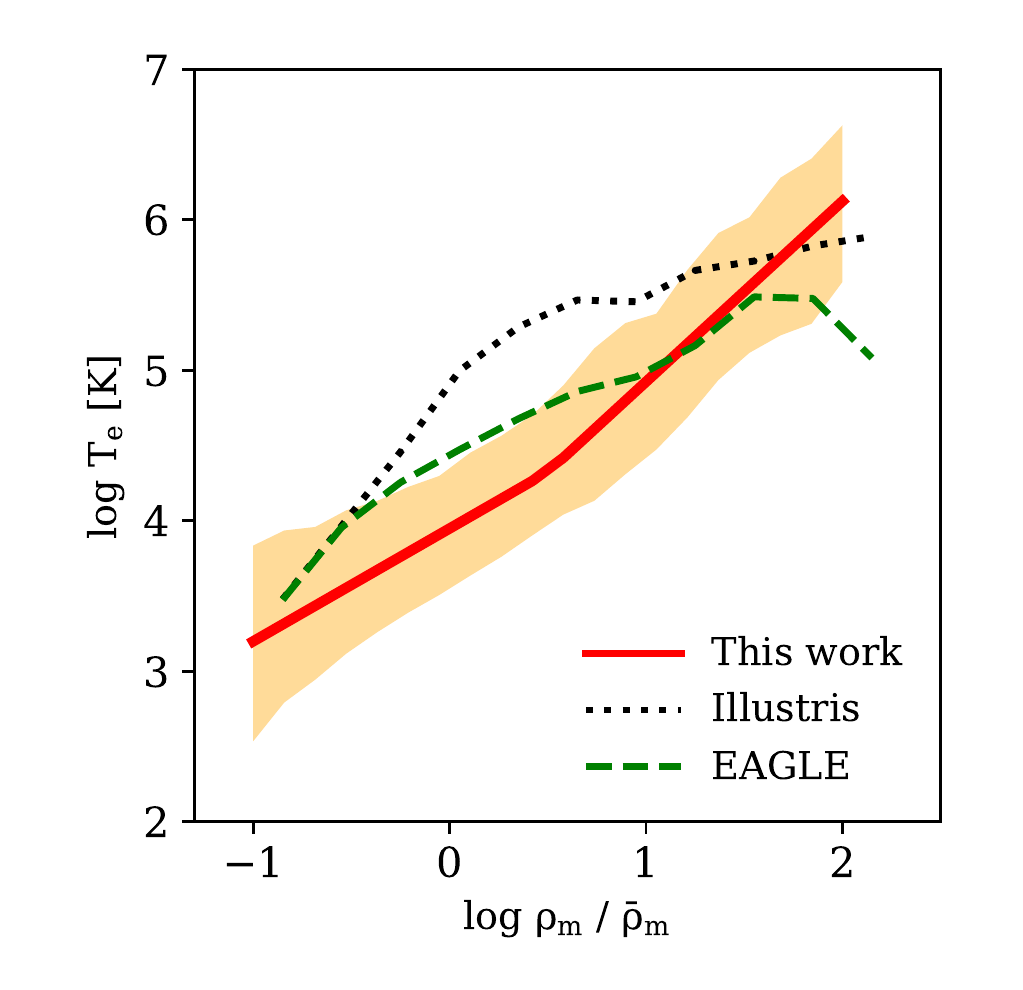}
\caption{The temperature - density relation derived 
assuming $\rho_{\rm B,ion}/\rho_{\rm m}\sim \Omega_{\rm B}/\Omega_{\rm m}=0.16$ and 
$n_{\rm e} = \rho_{\rm B,ion}\cdot [(1+f_{\rm H})/2m_{\rm p}]$, 
where $\rho_{\rm B,ion}$ is the ionized gas density, 
$f_{\rm H}=0.76$ is the hydrogen mass fraction and $m_{\rm p}$ is the 
proton mass. The red line shows the mean relation obtained with our method, with 
the orange band showing the $1\sigma$ dispersion estimated from the 
uncertainties in the constrained parameters. The dotted and dashed lines 
show the mean relations from Illustris and EAGLE, respectively. }
\label{fig_T_rho}
\end{figure}

Assuming that the ionized gas mass fraction with respect to the 
total mass within the grid-cells approximately equals the 
cosmic mean baryon fraction, as is motivated by the 
simulation results shown in Fig.\,\ref{fig_fB_ion},
one can convert the observed pressure - density relation to 
a relation between gas temperature and mass density. 
Specifically, we assume 
$\rho_{\rm B,ion}/\rho_{\rm m}\sim \Omega_{\rm B}/\Omega_{\rm m}=0.16$ and 
$n_{\rm e} = \rho_{\rm B,ion}\cdot [(1+f_{\rm H})/2m_{\rm p}]$, 
where $\rho_{\rm B,ion}$ is the ionized gas density, 
$f_{\rm H}=0.76$ is the hydrogen mass 
fraction and $m_{\rm p}$ is the proton mass, to obtain the 
electron temperature. The mean temperature - density relation thus 
derived for the entire observational sample is shown in 
Fig.\,\ref{fig_T_rho} by the solid line, along with the $1\sigma$ 
dispersion estimated from the uncertainties in the constrained 
parameters. As one can see, the average temperature is about 
$10^4\,{\rm K}$ in regions of mean density, 
$\rho_{\rm m} \sim {\overline\rho}_{\rm m}$,
increasing to $\sim 10^5\,{\rm K}$ for 
$\rho_{\rm m} \sim 10\,{\overline\rho}_{\rm m}$, 
the typical density for cosmic filaments and sheets
\citep[e.g.][]{ShenAbelMoSheth06}, and to $>10^{6}\,{\rm K}$ for  
$\rho_{\rm m} \sim 100\,{\overline\rho}_{\rm m}$, 
the typical density of dark matter halos. 

\subsection{Dependence on local tidal field} 
\label{ssec_DLTF}

\begin{figure*}
\includegraphics[width=0.85\linewidth]{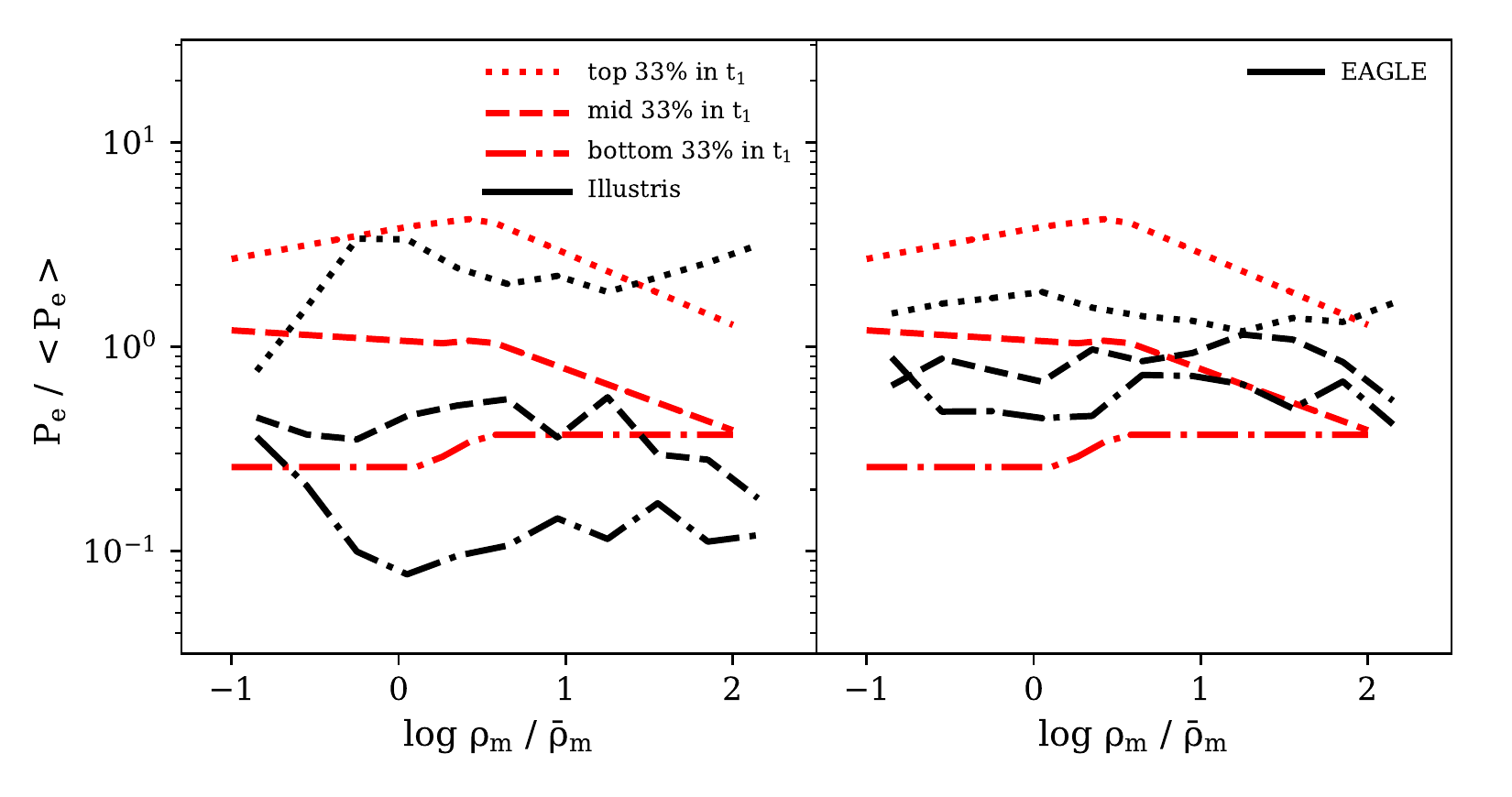}
\caption{The pressure - density relations obtained for the three
sub-samples of the grid-cells according to their ranking in the tidal 
field strength, $t_1$, at given density. 
The red lines in both panels show the mean relations obtained 
from the data, which are compared with the results of Illustris 
in the left panel and that of EAGLE in the right panel.   
For each case, the results are normalized by the corresponding 
mean relation shown in Fig.\,\ref{fig_P_rho}.} 
\label{fig_tidal}
\end{figure*}

To examine how the pressure - density relation depends 
on large-scale environment, we divide the grid-cells into 
three sub-samples, each containing a third of the total 
number of cells at the density in question, according to 
the ranking in the halo tidal field strength, $t_1$. 
We then constrain the pressure - density relations for the 
three sub-samples jointly, assuming different 
sets of the parameters for each of the sub-samples (see 
Table\,\ref{tab_param} for the parameters). 
The results are shown in Fig.\,\ref{fig_tidal}, 
with the pressure - density relation for each sample 
normalized by the mean relation shown in Fig.\,\ref{fig_P_rho}.
It is clear that, for a given density, the electron 
pressure is higher in regions of stronger tidal field,  
but the shapes of the relations are not very different
among the sub-samples. This result is consistent with 
that obtained from gas simulations, as we will see
in \S\ref{ssec_comparison}. The two breaking points 
in the relations obtained from the data arise because the relations
for the sub-samples have different values of $\rho_{\rm m,0}$ from 
that for the total sample. 

As mentioned in \S\ref{ssec_tidal}, we have tested a number 
of environmental quantities other than the tidal field 
as the second parameter that affects the thermal contents
of the IGM.  Among them, the distance of a grid-cell to the 
nearest massive halo is found to be nearly as a good 
indicator as the tidal field both from the simulation 
and from the application of our method to the observation.
This is not surprising, as the local tidal field is strongly 
correlated with the presence of massive structures nearby. 

\subsection{Comparisons with simulations}
\label{ssec_comparison}

In this subsection, we make comparisons of our results with 
gas simulations. Fig.\,\ref{fig_P_rho} shows that 
the pressure - density relation obtained from the data roughly 
matches those given by the simulations. The result obtained 
from shuffling the grid-cells is also found to be very similar 
to the total average pressure from the simulations, as 
indicated by the horizontal dashed line in Fig.\,\ref{fig_P_rho}. 
In more detail, the observed pressure - density relation 
matches well with that given by EAGLE in both amplitude and shape, 
but Illustris significantly over-predicts the gas pressure in the 
intermediate density range from 
$\sim {\overline\rho}_{\rm m}$ to $\sim 10\,{\overline\rho}_{\rm m}$. 
As mentioned earlier, the difference between Illustris and 
EAGLE is likely caused by the different implementations 
of feedback processes adopted in the two simulations. 
Indeed, as shown in \citet{vogelsberger14}, the strong AGN feedback 
adopted in Illustris can heat a significant fraction of the IGM 
at large distances from over-dense regions, 
which may explain the high pressure seen in the 
intermediate density range. Our results, however, suggest that 
such strong feedback may not be favored by the data. 

The dotted and dashed lines in Fig.\,\ref{fig_T_rho} 
are the average temperatures obtained directly from Illustris and EAGLE, 
respectively. Thus, if the ionized baryon fraction is approximately 
equal to the universal baryon fraction on scales of 
$\sim 1\,h^{-1}{\rm Mpc}$ in the real universe, as is assumed 
in deriving the temperature - density relation from the data,   
then the average IGM temperature predicted by Illustris 
in the density range $(1-10){\overline\rho}_{\rm m}$ is 
too high. The prediction of EAGLE is consistent with 
the observation, given the uncertainty in the data. 
This again shows that the data are already capable of 
providing interesting constraints on models of galaxy formation.

Finally, in Fig.\,\ref{fig_tidal}, we compare observation 
and simulation results in their dependence on the strength 
of local tidal field. Here the simulation results are 
normalized by the corresponding mean relations shown 
in Fig.\,\ref{fig_P_rho}. The simulation results are consistent 
with the observational data in that the gas pressure at a given mass 
density is higher in regions of stronger tidal field.
We have checked that the average ionized gas fraction 
is quite independent of the tidal strength in the simulations, 
and so the higher pressure in stronger tidal field is due to 
higher gas temperature rather than higher gas density. 
The dependence on the tidal field predicted by Illustris
is much stronger than that by EAGLE. If shock heating 
by gravitational collapse has similar effects  
in both simulations, the difference in the tidal field 
dependence should then be a result of the different 
prescriptions of feedback used in the two simulations.  
The feedback effects on the IGM are expected to depend
strongly on the local tidal field. For example,  
a cell with relatively low density but high $t_1$
must have some massive structures nearby to produce 
the strong tidal field. Such massive structures are 
also where strong stellar/AGN feedback is produced.  
In fact, as can be seen from 
figure 4 in \citet{vogelsberger14}, feedback effects
on the IGM are clearly more important in the neighborhoods 
of more massive structures, where the tidal field is also 
stronger. It is interesting to note that, 
in both simulations, the strongest tidal field 
dependence occurs in regions with 
$\rho_{\rm m}\sim {\overline\rho}_{\rm m}$, although the 
signal is rather weak, indicating that the gas temperatures 
in such regions may be affected the most by feedback from 
nearby structures. The tidal field 
dependence obtained from the data is weaker than that predicted
by Illustris but stronger than that by EAGLE. We have 
estimated the velocity dispersion of dark matter 
particles, $\sigma$, in individual grid cells in the 
simulations, and examined the average of $\sigma^2$ 
for cells of different $\rho_{\rm m}$ in regions of 
different tidal strengths. We found that the dependence 
of $\sigma^2$ on the tidal strength in the 
intermediate density range, $\rho_{\rm m}\sim (1-10)
{\overline\rho}_{\rm m}$, is weaker than that of gas 
temperature, both in the observation and particularly in 
Illustris. If we take $\sigma^2$ as a measure of heating 
by gravitational collapse, then non-gravitational 
processes, such as stellar and AGN feedback, must have 
played an important role in heating the IGM in the 
intermediate density range. 

\section[summary]{SUMMARY AND DISCUSSION}
\label{sec_sum}

In this paper we examine the thermal energy contents of the  
IGM over three orders of magnitude in both mass density and gas 
temperature using thermal Sunyaev-Zel'dovich effect (tSZE). 
Our results are based on {\it Planck} tSZE map and the cosmic density 
field, reconstructed for the SDSS DR7 volume and sampled 
on a grid of cubic cells of $(1h^{-1}{\rm Mpc})^3$,  
together with a matched filter technique employed to 
maximize the signal-to-noise. 

Our results obtained by matching all the grid cells show that the 
pressure - density relation of the IGM is roughly a power law given 
by an adiabatic equation of state, with some indication of 
a steepening at densities higher than about $10$ times the mean 
density of the universe. The result from shuffling the grid-cells 
shows a nearly zero slope for the pressure-density relation, 
demonstrating that the relation obtained by our method indeed 
captures the thermal properties of the gas that produces the 
observed tSZE. 

Using the simulation result 
that the ionized gas mass fractions within individual 
grid cells are about equal to the universal baryon 
fraction, we convert the pressure - density relation to 
a temperature - density relation. The result shows that
the average temperature is about $10^4\,{\rm K}$ in regions 
of mean density, $\rho_{\rm m} \sim {\overline\rho}_{\rm m}$,
increasing to about $10^5\,{\rm K}$ for 
$\rho_{\rm m} \sim 10\,{\overline\rho}_{\rm m}$,
the typical density for cosmic filaments and sheets, and to 
$>10^{6}\,{\rm K}$ for  $\rho_{\rm m} \sim 100\,{\overline\rho}_{\rm m}$, 
the typical density of virialized dark matter halos.

The thermal energy content of the IGM is also found to be higher 
in regions of stronger tidal fields. By dividing grid cells 
into three equal-sized sub-samples according to the local tidal 
field strength, we find that the average gas temperature in the 
sub-sample of highest tidal field is a factor of $10$ higher than 
that in the lowest tidal field sub-sample. Such an increase of
temperature in intermediate density regions is stronger by a factor 
of two than that expected from the increase of average 
velocity dispersion of dark matter in simulations, suggesting 
that feedback from galaxy formation may be responsible for 
the increase in gas temperature. 

We compare our results with those obtained from two hydrodynamic 
simulations, Illustris and EAGLE. While the simulations can 
reproduce the general trends observed in the observation, 
such as the increases of gas pressure and temperature with 
dark matter density and the strength of local tidal field, 
there are significant discrepancies between the two simulations, 
as well as between the simulations and our observational 
results. Within the uncertainties of the data, the predictions 
of EAGLE are consistent with the data. However, Illustris predicts 
significantly higher gas pressure and temperature in the 
intermediate density range, $\rho_{\rm m} \sim (1 - 10) 
{\overline\rho}_{\rm m}$, than both the observation
and EAGLE. The dependence on tidal field strength
predicted by Illustris is also too strong in comparison 
with the observational data and EAGLE. We suspect that 
these differences are produced by the strong AGN feedback adopted 
in Illustris that can heat the IGM at large distances from 
massive structures. 

Our results clearly demonstrate the promise of using SZE, 
combined with reconstructed density field, to study both the IGM and 
the galaxy formation processes that produce them. 
This approach is complementary to absorption line studies, 
in that it is not constrained by a limited number of 
lines of sight, and that corrections for metallicity and 
ionization effects are not needed to obtain the total gas 
mass. It also complements X-ray observations, in that 
it is more sensitive to the diffuse warm-hot gas that is expected  
to dominate the IGM. In the future, when high-resolution SZE
data are available, the same approach as developed here
can be used to study not only the detailed distribution and 
state of the IGM, but also to investigate how the IGM is 
related to and affected by galaxies and AGNs in the cosmic web.

\section*{ACKNOWLEDGEMENTS}

HJM acknowledges the support from NSF AST-1517528. This work is also 
supported by the China 973 Program (No. 2015CB857002) and the National Science 
Foundation of China (grant Nos. 11233005, 11621303, 11522324, 
11421303, 11503065, 11673015, 11733004). 
We thank the Planck collaboration for making the full-sky maps public, and  
the Virgo Consortium for making the EAGLE simulation data available. 
The EAGLE simulations were performed using the DiRAC-2 facility at Durham, 
managed by the ICC, and the PRACE facility Curie based in France at TGCC, 
CEA, Bruy\`eresle-Ch\^atel.


\label{lastpage}

\end{document}